\documentclass[a4paper,11pt]{article}
\usepackage{jinstpub}
\usepackage{lineno}

\title{\boldmath Performance of the Mass Testing Setup for Arrays of Silicon Photomultipliers in the TAO Experiment.}

\author[a,1]{A. Rybnikov,\note{Corresponding author.}}
\author[a]{N. Anfimov,}
\author[b]{M. Qu,}
\author[a]{A. Chetverikov,}
\author[c,d]{G. Cao,}
\author[a]{D. Fedoseev,}
\author[c]{H. Wang,}
\author[a]{V. Kozhukalov,}
\author[a]{V. Sharov,}
\author[a]{S. Sokolov,}
\author[a]{A. Sotnikov}

\affiliation[a]{Joint Institute for Nuclear Research,\\
6 Joliot-curie str., Dubna, Moscow Region, Russia, 141980}
\affiliation[b]{Zhengzhou University,\\
No.100 Science Avenue, Zhengzhou City, Henan Province, China, 450001}
\affiliation[c]{Institute for High Energy Physics,\\
19B Yuquan Road, Shijingshan District, Beijing, China, 100049}
\affiliation[d]{School of Physical Sciences, University of the Chinese Academy of Sciences,\\
Beijing, China, 101408}

\emailAdd{arv@jinr.ru}

\abstract{
Modern neutrino physics detectors often employ thousands, and sometimes even hundreds of thousands, of Silicon Photomultipliers (SiPMs). The TAO experiment \cite{junotaocdr} is a notable example that utilizes a spherical scintillator barrel with a diameter of 1.8 meters, housing approximately 130,000 SiPMs organized into 4,100 tiles.  Each tile with size of $5\times5~$cm$^2$ consists of a 32-SiPM array functioning as a single detector unit. To achieve an unparalleled energy resolution of 2\% at 1 MeV within this volume, the SiPMs must possess cutting-edge parameters, including a photon detection efficiency (PDE) exceeding 50\%, cross-talk of approximately 10\%, and an extremely low dark count rate (DCR) below $50~$Hz/mm$^2$. Maintaining the setup at a negative temperature of $-50^\circ$C is necessary to achieve the desired DCR.
This article presents the setup and methods employed to individually characterize the mass of SiPMs across all 4,100 tiles at the specified negative temperature.}

\keywords{Photon detectors for UV, visible and IR photons (solid-state) (PIN diodes, APDs, Si-PMTs, G-APDs, CCDs, EBCCDs, EMCCDs, CMOS imagers, etc), Neutrino detectors}

\begin{document}
\maketitle
\flushbottom

\section{Introduction}
\label{sec:intro}
A silicon photomultiplier (SiPM) is an array of small avalanche photodiodes called pixels that operate in Geiger mode above the breakdown voltage.
This design enables the detection and handling of very low-intensity light in a ''proportional'' mode within a range determined by the total number of pixels.
SiPMs are extensively used in diverse areas, including experimental physics and medical diagnostic equipment. They offer several advantages over conventional photomultipliers, such as high detection efficiency, insensitivity to magnetic fields, compact size, and a comparable gain of up to 10$^6$. However, SiPMs also have some drawbacks, including a high dark count rate, optical crosstalk, and temperature-dependent parameters.

Currently, a new experiment called the TAO detector \cite{junotaocdr} is in the assembling stage  as a complement to the JUNO neutrino project \cite{JUNO}. The purpose of the TAO is to accurately measure the spectrum of primary anti-neutrinos originating from nuclear reactors. It consists of a spherical vessel filled with liquid scintillator and equipped with photosensors.

In order to achieve the required energy resolution of less than 2\% at 1 MeV for the TAO experiment, it is crucial to maximize the light detection efficiency and minimize noise levels. Silicon photomultipliers (SiPMs) are considered a promising choice for the TAO detector due to their high photon detection efficiency (PDE). To mitigate the high noise levels associated with SiPMs, it is proposed to operate the TAO detector at a low temperature of approximately $-50~^\circ$C~\cite{tao2}.

TAO experiment will utilize 4,100 tiles, each measuring 5x5 cm², composed of 32 pieces of 12x6 mm² Hamamatsu SiPMs. The SiPMs on each tile are grouped into 16 SiPM channels, with each channel consisting of two parallelly connected SiPMs\footnote{Hereinafter we will refer to it as SiPM}. Each individual SiPM should undergo meticulous characterization to determine its photon detection efficiency (PDE), gain, cross-talk, and dark count rate (DCR). This thorough characterization process is crucial to ensure optimal performance of the detector and to achieve uniformity of these parameters. By minimizing variations in these characteristics, the impact on the constant term of the energy resolution can be reduced.

\section{Mass-Testing Setup}
\label{sec:setup}
\begin{figure}[htbp]
\centering
\includegraphics[width=.4\textwidth]{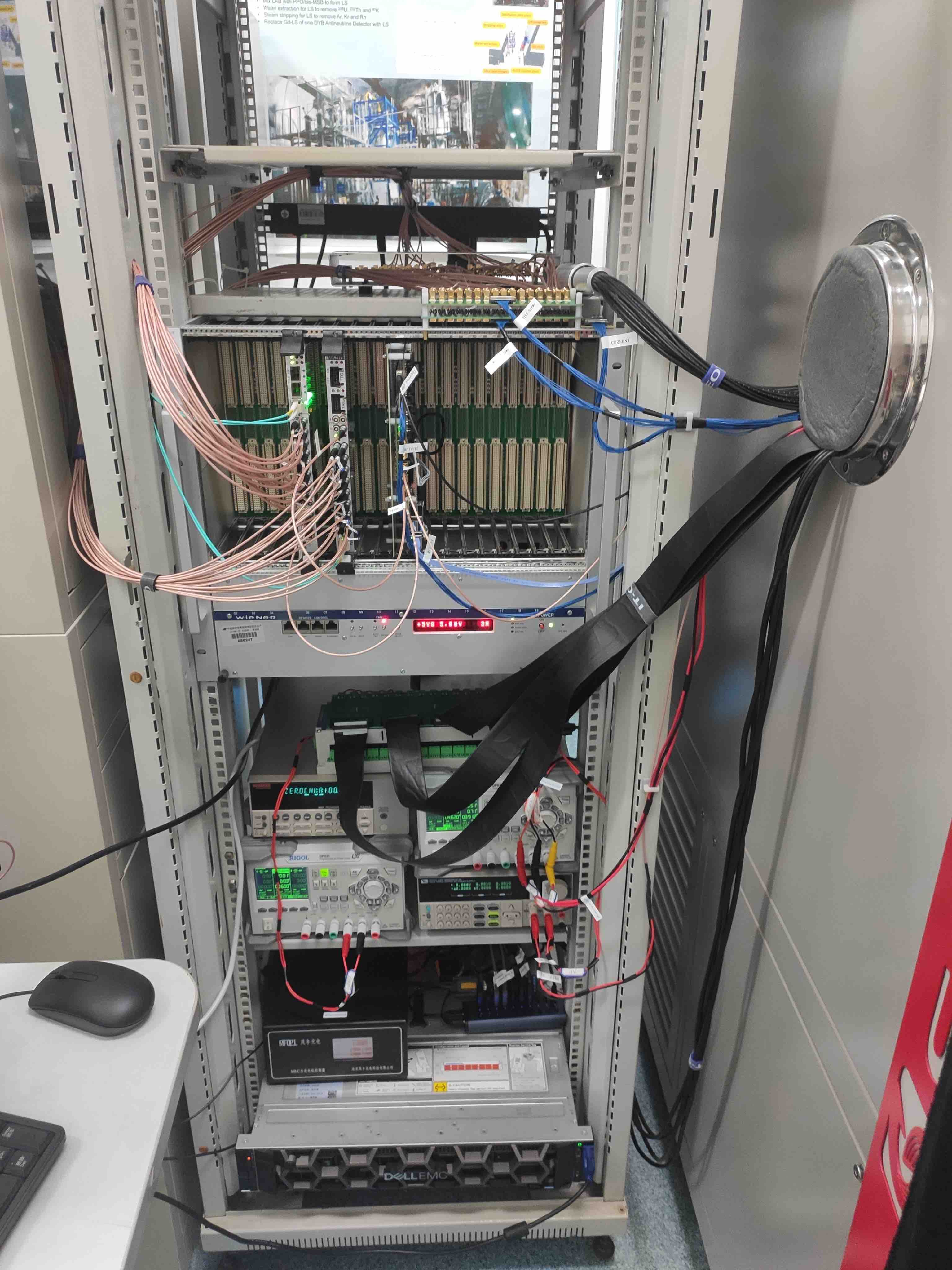}
\qquad
\includegraphics[width=.4\textwidth]{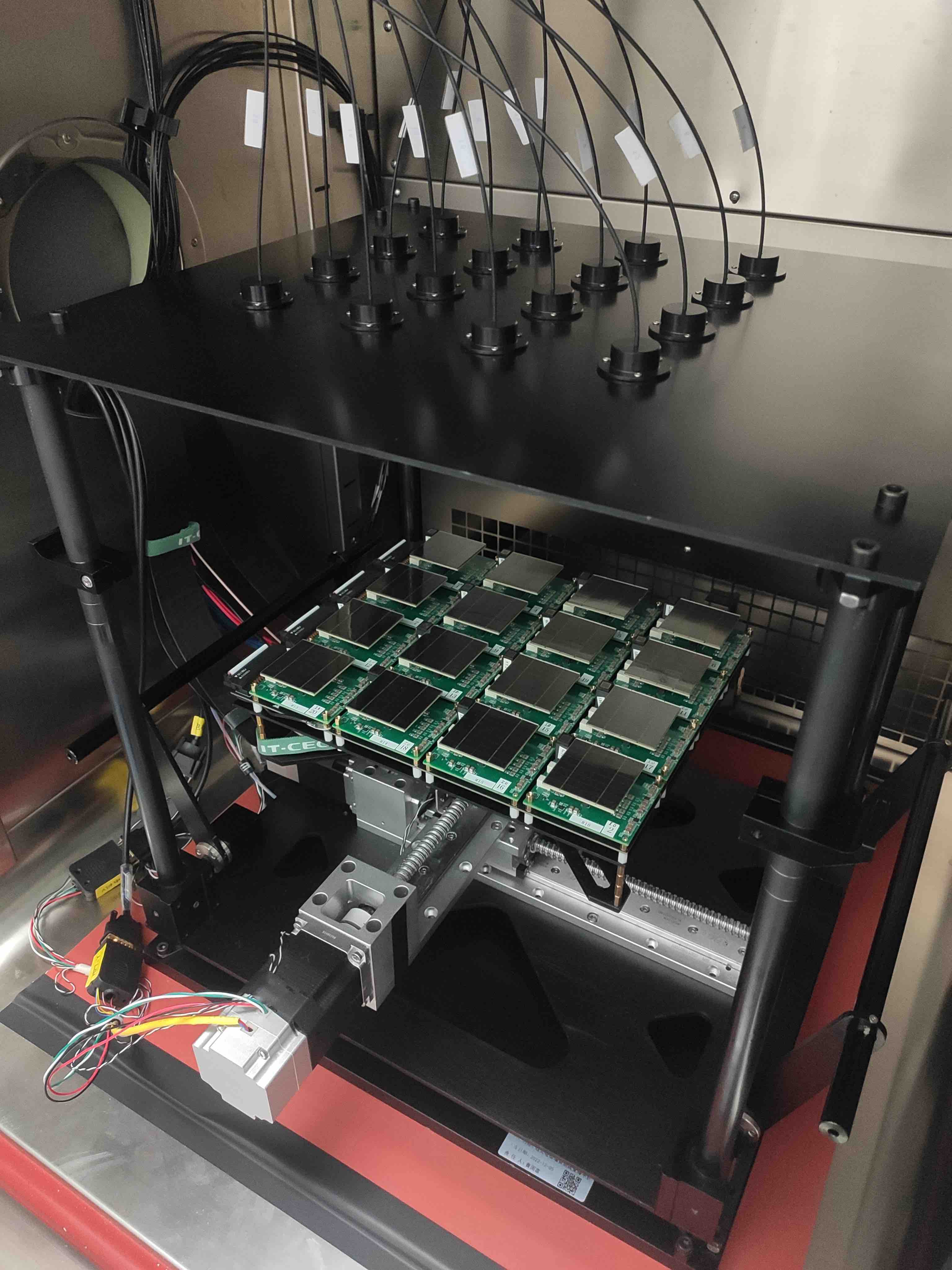}
\caption{Mass-testing setup for SiPM tiles. \label{fig:setup}}
\end{figure}

\begin{figure}[htb]
\begin{center}  
\includegraphics[width=1.0\linewidth]{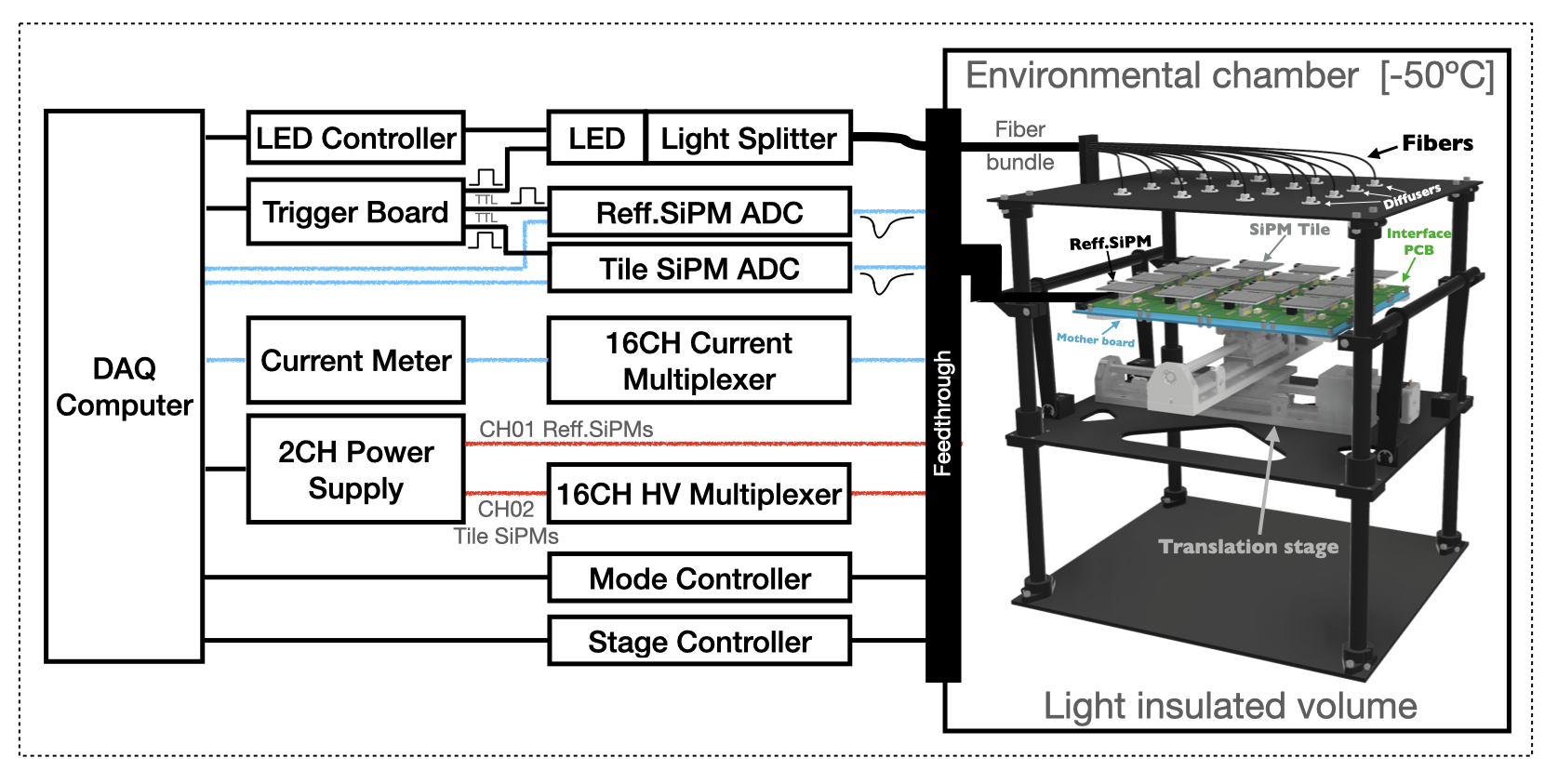} 
\caption{Scheme of the mass-testing setup}
\label{fig:setup-scheme} 
\end{center}
\end{figure}

The testing setup (see  figure~\ref{fig:setup} and 
 \ref{fig:setup-scheme}) for characterizing the SiPMs consists of a PCB motherboard designed to accommodate 16 SiPM tiles. These tiles are positioned on the motherboard using 16 adapter boards, ensuring proper alignment and tile connector. Each adapter board is equipped with an operational amplifier to gain the response of the SiPMs before transmitting it through a cable line. In order to achieve uniform light illumination across the tiles, a fiber splitter, as described in \cite{optical-splitter}, is employed. For this purpose, 16 fibers with PTFE diffusers at their ends are utilized. These diffusers are placed on top of each tile to ensure the uniform distribution of light.

To provide a stable and consistent light intensity, an LED source \cite{hvsys} is employed. The LED source incorporates a PIN photodiode within a feedback loop, enabling precise control and stabilization of the light intensity. This setup guarantees that the SiPMs receive a consistent light input during the characterization process. 

To monitor the light intensity across the SiPMs, 16 individual reference SiPMs are placed adjacent to each tile, serving as monitors. These monitors allow for the assessment of the light field distribution and provide valuable information on any variations or irregularities in the illumination across the SiPMs. 

To facilitate a comprehensive scan of the light field, the entire motherboard is mounted on a translation stage equipped with step motors, enabling precise movement in two directions. This capability allows for systematic scans of the light field over the SiPMs, ensuring thorough characterization and assessment of their performance.

In addition to its functionality, the setup is designed with a convenient lifting mechanism. This allows the motherboard to be lifted up for illumination purposes and then lowered for the installation and removal of individual tiles. By lifting the motherboard, the SiPMs are exposed to the uniform illumination provided by the fiber splitter and LED source. Furthermore, when it comes to tile installation or disinstallation, lowering the motherboard provides a convenient working space. This allows for efficient handling and placement of the individual tiles onto the adapter boards.

For the testing of all SiPMs on the 16 tiles, a Keithley 6487/E picoammeter and ITECH IT6300 voltage source \cite{ITECH} are utilized, along with two 16-channel multiplexers. The multiplexers play a major role in enabling biasing and current measurements across the SiPMs.

The first multiplexer is employed for the readout of SiPMs in different positions across the tiles. The same voltage is applied from ITECH source to the SiPMs in a specific position on all 16 tiles. For example, all SiPMs at position number 1 on the 16 tiles receive the same applied voltage. By sequentially switching to another SiPM position across the tiles, the readout process covers all SiPMs on the 16 tiles. This allows for simultaneous readout of the SiPM responses using a 16-channel analog-to-digital converter (ADC) TQDC16VS\cite{AFI} with a high resolution of 14 bits and a sampling rate of 125 MS/s.

The second multiplexer is dedicated to current measurements. Since the Keithley 6487/E picoammeter is a single-channel instrument, switching between the 16 SiPMs on the 16 tiles is necessary. The multiplexer is responsible for selecting a specific tile, and then, using the other multiplexer, individual SiPMs on that particular tile are switched one by one. This enables the measurement of all SiPMs on the selected tile. The process is repeated as the multiplexer switches to another tile, ensuring  measurements across all SiPMs on each tile.

Another channel of the ITECH voltage source is dedicated to supplying power to 16 reference SiPMs. These SiPMs serve as calibration standards for the absolute measurement of the SiPMs on the test tiles. They are carefully calibrated in terms of their absolute photon detection efficiency (PDE), which allows them to serve as reliable references for evaluating the PDE of the SiPMs on the test tiles.

\section{Analysis Matter}
\label{sec:stats}
\begin{figure}[htbp]
\centering
\includegraphics[width=\textwidth]{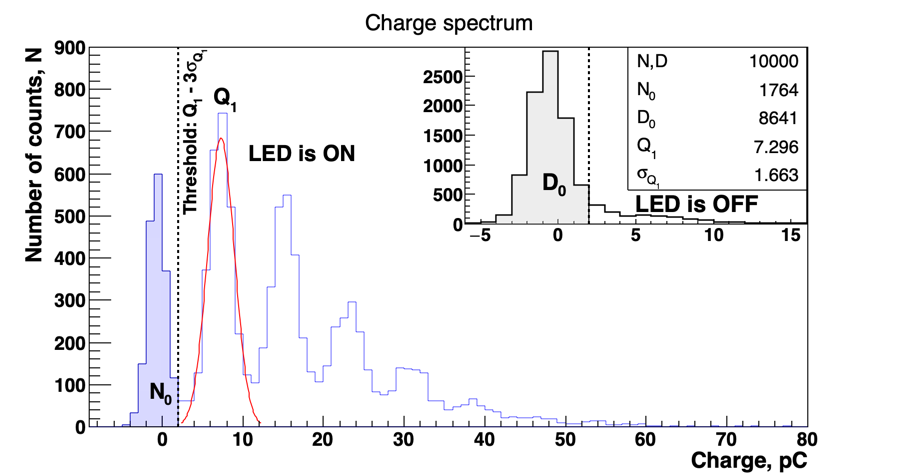}
\caption{Typical charge spectrum of SiPMs. Charge spectrum illustrates the pedestal method for evaluation of an average number of photoelectrons. Shaded area and $N_0$ - pedestal events on signal spectrum, shaded area and $D_0$ - pedestal events on dark spectrum (incorporated picture)  \label{fig:spectrum}}
\end{figure}

During the data acquisition process, each SiPM undergoes scanning to obtain the single photoelectron (p.e.) charge spectrum, as presented in figure~\ref{fig:spectrum}. This spectrum represents the distribution of charge signals generated by individual photoelectrons detected by the SiPM. By analyzing this spectrum and applying statistical methods, we can estimate main parameters as the average number of photoelectrons, denoted as $\mu$, the cross-talk value and the pixel gain. 

The pixel gain, denoted as $G_{pix}$, is estimated by evaluating the charge between two neighboring peaks, denoted as $Q_1$, e.g. single electron peak  and the pedestal peak in the spectrum, expressed in terms of the number of electron charges $q_e$. However, to obtain the actual pixel gain, the measured charges need to be normalized by the total gain factor of all the amplifiers $K_{amp}$ used in the measurement setup. The expression is as follows

\begin{equation}
    G_{pix} = \frac{Q_1}{K_{amp} \cdot q_e}.
\end{equation}

The pedestal is the SiPM response without any incident photons. The number of events in the pedestal measurement is denoted as $N_0$, while the total number of events as $N$. To estimate $\mu$, we utilize the Poisson distribution and make use of the statistics obtained from the pedestal measurements.

In order to account for the noise present in the SiPMs, we acquire a dark spectrum without any incident light.  By analyzing this dark spectrum and utilizing the same statistical methods, we can estimate the average number of photoelectrons, $\mu$, while considering the presence of noise \cite{optimization}

\begin{equation}
\mu = -\ln\left(\frac{N_0}{N}\frac{D}{D_0}\right).
\end{equation}
where $D_0$ - pedestal in the dark spectrum and $D$ - the total number of events in the dark spectrum. 

To produce the dark spectrum, we analyze the same dataset of waveforms in the region where the LED signal is absent. In this region, the SiPM response is solely due to noise. Therefore, the total number of events in the dark spectrum, D, is the same as the total number of events in the spectrum with the LED signal, N.
Considering this, we can simplify the equation for estimating $\mu$ as follows

\begin{equation}
\label{eq:est_mu}
\mu = -\ln\left(\frac{N_0}{D_0}\right).
\end{equation}

The statistical accuracy in this case is indicated by the uncertainty or error associated with the estimated average number of photoelectrons as \cite{optimization}\footnote{In equation (36) of the reference \cite{optimization}, we can substitute  $e^{\xi_0}$ with $N/N_0$ and $e^{\lambda_0}$ with $D/D_0$ and neglect 2.}

\begin{equation}
    \label{eq:dispersion_mu_00_noise}
      \sigma_{\mu} \approx \sqrt{\frac{D_0 + N_0}{D_0N_0}}.
\end{equation}

In the study mentioned in \cite{optimization}, it was indicated that the best statistical accuracy was achieved around $\mu\approx1.6$. Without the presence of noise, the statistical accuracy was estimated to be around 1.24\% with a total number of events $N=10,000$. To ensure a statistical accuracy of less than 1\% with the presence of noise and in a broader range of $\mu$ from 1 to 2 p.e., a higher number of events is required. To achieve this, we set $N=30,000$.

The cross-talk parameter, which represents the number of fired pixels in the first generation and is often denoted as $\lambda$, signifies the probability of a photoelectron-induced signal in a neighboring pixel. To estimate the cross-talk value in SiPMs, the Generalized Poisson distribution is utilized, as described in reference \cite{vinogradov}

\begin{equation}
    \label{eq:cross-talk_GP}
 P(n|\mu,\lambda) = \frac{\mu(\mu+n\lambda)^{n-1}e^{-(\mu+n\lambda)}}{n!}
\end{equation}

The cross-talk parameter $\lambda$ can be estimated from the mean value in the histogram by subtracting the pedestal value,  which represents the average SiPM response, denoted as $S$.  The estimation of $\lambda$ is done as follows\footnote{In equation for the mean in the table 2 of the reference \cite{vinogradov}, we can substitute  $E[x]$ with $S/Q_1$}

\begin{equation}
 \label{eq:cross-talk}
    \lambda = 1 - \frac{\mu Q_1}{S}
\end{equation}

The errors for $Q_1$ and $S$ are determined based on statistical analysis of the data, specifically through Gaussian fitting for $\delta_{Q_1}$  and analysis of the mean error in histogram for $\delta_{S}$ accounting the pedestal error $\delta_{Q_0}$. The uncertainty for $\lambda$ is obtained by propagating the errors $\delta_{\mu}$\footnote{We calculate errors at the 68\% confidence level, thus we utilize the standard deviation $\sigma_{\mu}$ from \eqref{eq:dispersion_mu_00_noise} to estimate the uncertainty.}, $\delta_{Q_1}$, and $\delta_{S}$ as

 \begin{equation}
  \label{eq:cross-talk_error}
    \delta_{\lambda} = \frac{1}{S}\sqrt{(Q_1\delta_{\mu})^2 +(\mu\delta_{Q_1})^2 + (\frac{Q_1 \mu}{S}\delta_{S})^2}
\end{equation}

\begin{figure}[htbp]
\begin{center}  
\includegraphics[width=1.0\linewidth]{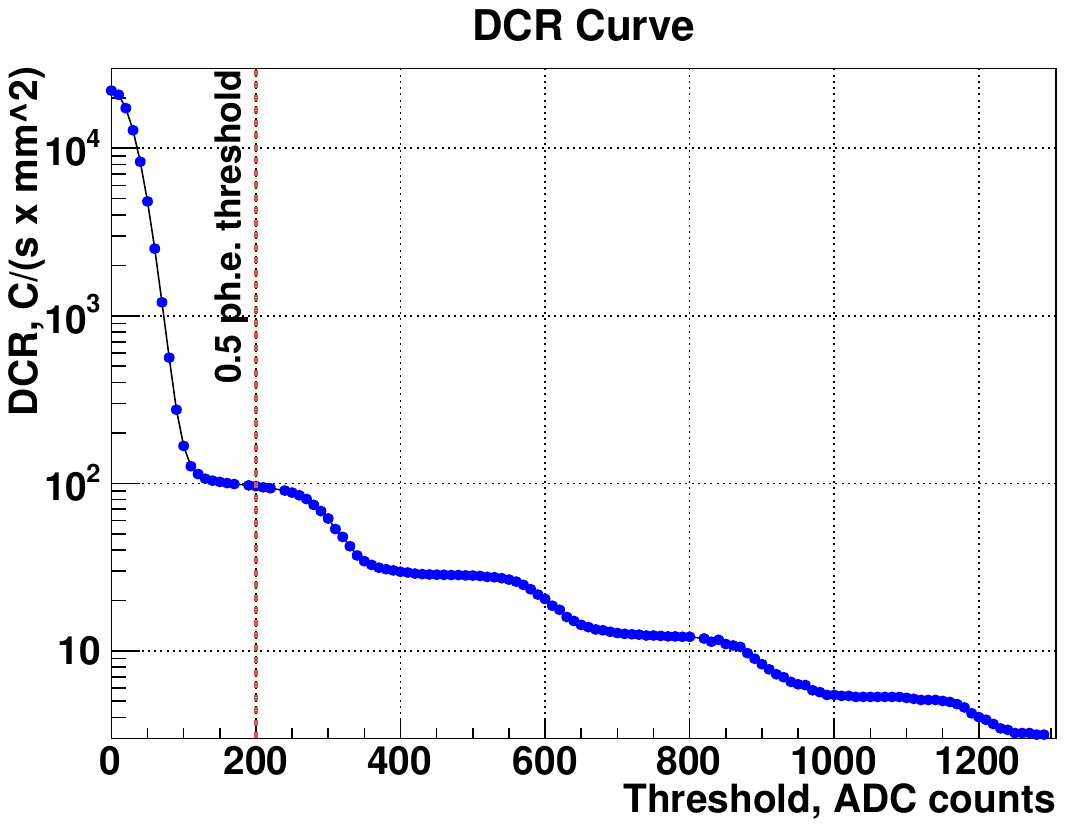} 
\caption{An example of the software-reconstructed dark count rate curve for a 5V overvoltage.}
\label{fig:dcr-curve} 
\end{center}
\end{figure}

The dark count rate (DCR) is estimated through waveform analysis.  
This analysis utilizes the same data collected during the main run but focuses on a window denoted as $W$ of approximately 10 microseconds preceding the signal from the light source.
The time window is reduced by half of the pulse width on both sides to ensure the accurate counting of signals at the edges of the frame.
To mitigate the inclusion of high-frequency noise, a moving average filter is applied.
For each specific oscillogram, an algorithm is employed to identify signal instances above a predefined threshold within the selected time range. 
The counts denoted as $C$ of these instances are summed when processing all the $N$ oscillograms in the data file. 
The result is then normalized to  1 second and the total area $S=12\times12~$mm$^2$ of the SiPM, and it is represented in units of Hz per square millimeter as follows

\begin{equation}
 \label{eq:DCR}
    DCR = \frac{C}{W \times N \times S},
\end{equation}
and its error
\begin{equation}
 \label{eq:DCR_error}
    \sigma_{DCR} = \frac{\sqrt{C}}{W \times N \times S}.
\end{equation}
The equation \ref{eq:DCR_error} represents a statistical error following a Poisson distribution of counts $C$.
By obtaining the DCR at various thresholds, we construct the DCR curve (see fig.~\ref{fig:dcr-curve}) and determine the appropriate threshold position in accordance with the mass-testing requirements (typically set at 0.5 photoelectrons). The DCR value at this threshold is used to estimate the DCR level for a specific SiPM

\section{Software}
\label{sec:soft}

\begin{figure}[htb]
\begin{center}  
\includegraphics[width=1.0\linewidth]{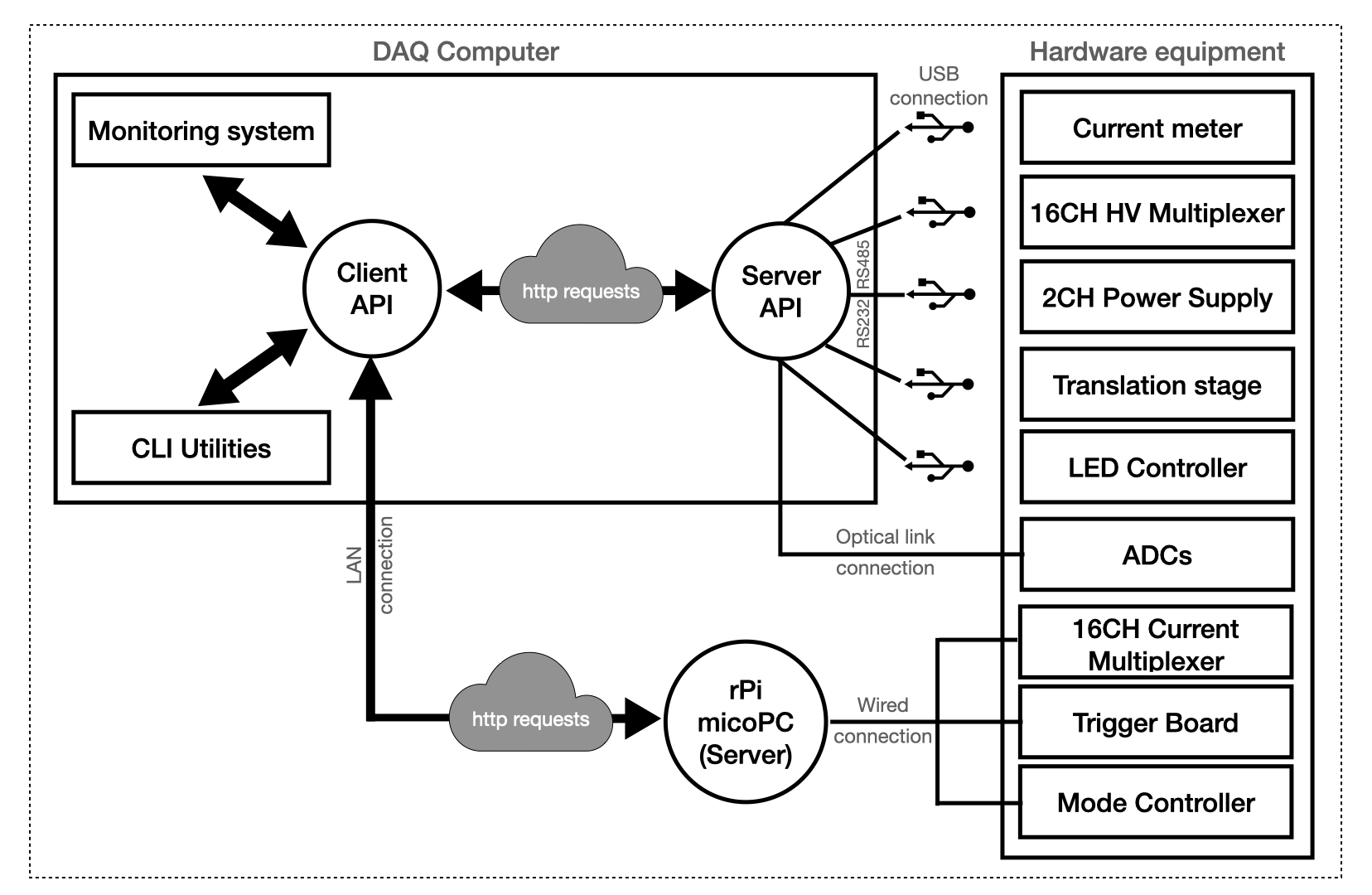} 
\caption{Architecture of DAQ software for the mass-testing setup}
\label{fig:daqsoft} 
\end{center}
\end{figure}

The scanning process is facilitated through a control software that is installed on the DAQ (Data Acquisition) computer. This software, depicted in Figure~\ref{fig:daqsoft}, follows a client-server approach in its implementation. Each hardware component of the mass-testing setup software consists of both a server and a client part. The server runs as a system service, providing an interface to manage the hardware, while the client offers a set of commands for equipment control via command-line or graphical interface integration with a monitoring system. This architecture enables efficient communication and coordination, streamlining the SiPM characterization process.

The server segment of the software is modular, comprising multiple components that manage distinct hardware elements of the mass-testing setup. Most of these hardware elements connect to the DAQ machine via USB. For slow control and monitoring, we utilize RS-232 and RS-485 protocols.

The ADC board interfaces with the DAQ machine through an optical 10GBit Ethernet connection. During data acquisition, the typical event rate is approximately 600Hz.

The ADC server software has two distinct components. Each implements its proprietary protocol operating over UDP: one for data acquisition and the other for ADC board management. Both data and management traffic traverse the same optical interface.

\begin{figure}[htb]
\begin{center}  
\includegraphics[width=1.0\linewidth]{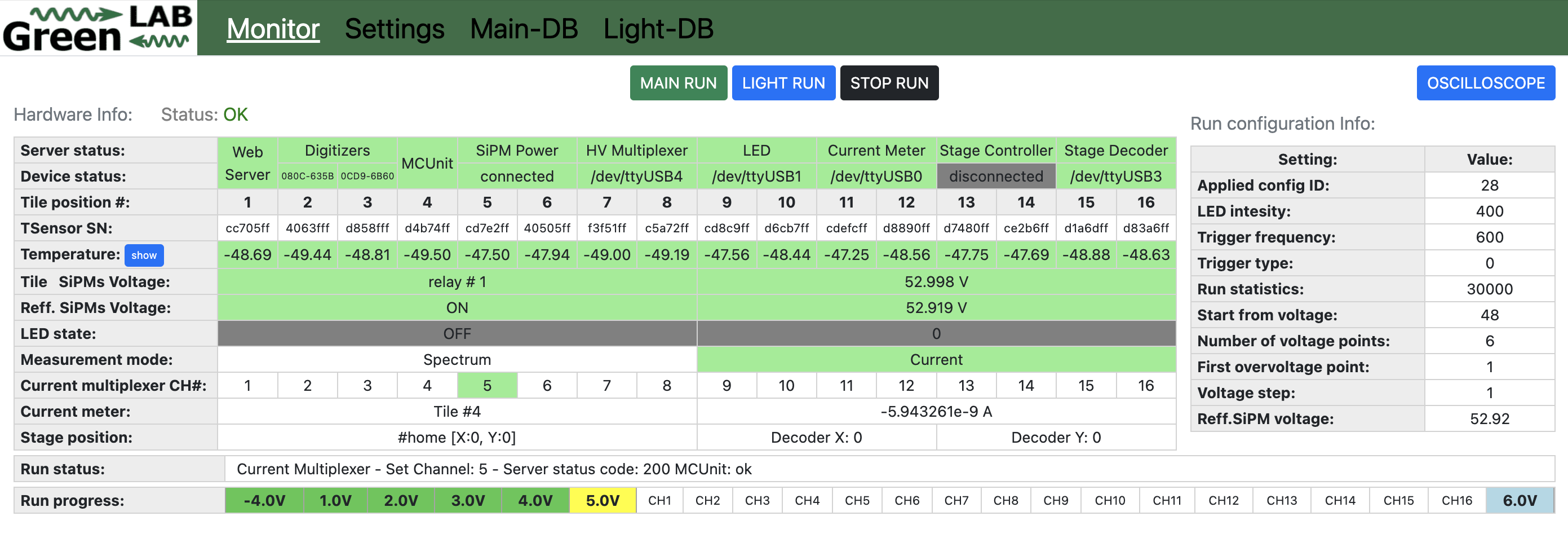} 
\caption{Monitoring system of the mass-testing setup}
\label{fig:monitor} 
\end{center}
\end{figure}

The monitoring system, shown in Figure~\ref{fig:monitor}, provides real-time monitoring of the scanning process and alerts the user to any hardware malfunctions. It offers a user-friendly interface for initiating two types of runs: light field scanning and data acquisition for SiPM tiles. These runs can be combined to obtain comprehensive characteristics of the SiPM tiles. The monitoring system provides efficient feedback on the LED intensity.

During the joint run, certain data such as current measurements and temperature readings are recorded and stored in the light and main databases, respectively. Another portion of the dataset, specifically the oscillograms, is digitized using ADCs, saved to the disk, and subsequently transferred to the IHEP cluster. Once transferred to the cluster, the data undergoes analysis using the analysis software on the supercomputing system. The analysis software extracts the required parameters from the data files and populates the database with the extracted information, ensuring comprehensive and organized storage of the SiPM characterization data.

\section{Setup performance}
\label{sec:perform}
\begin{figure}[htb]
\begin{center}  
\includegraphics[width=1.0\linewidth]{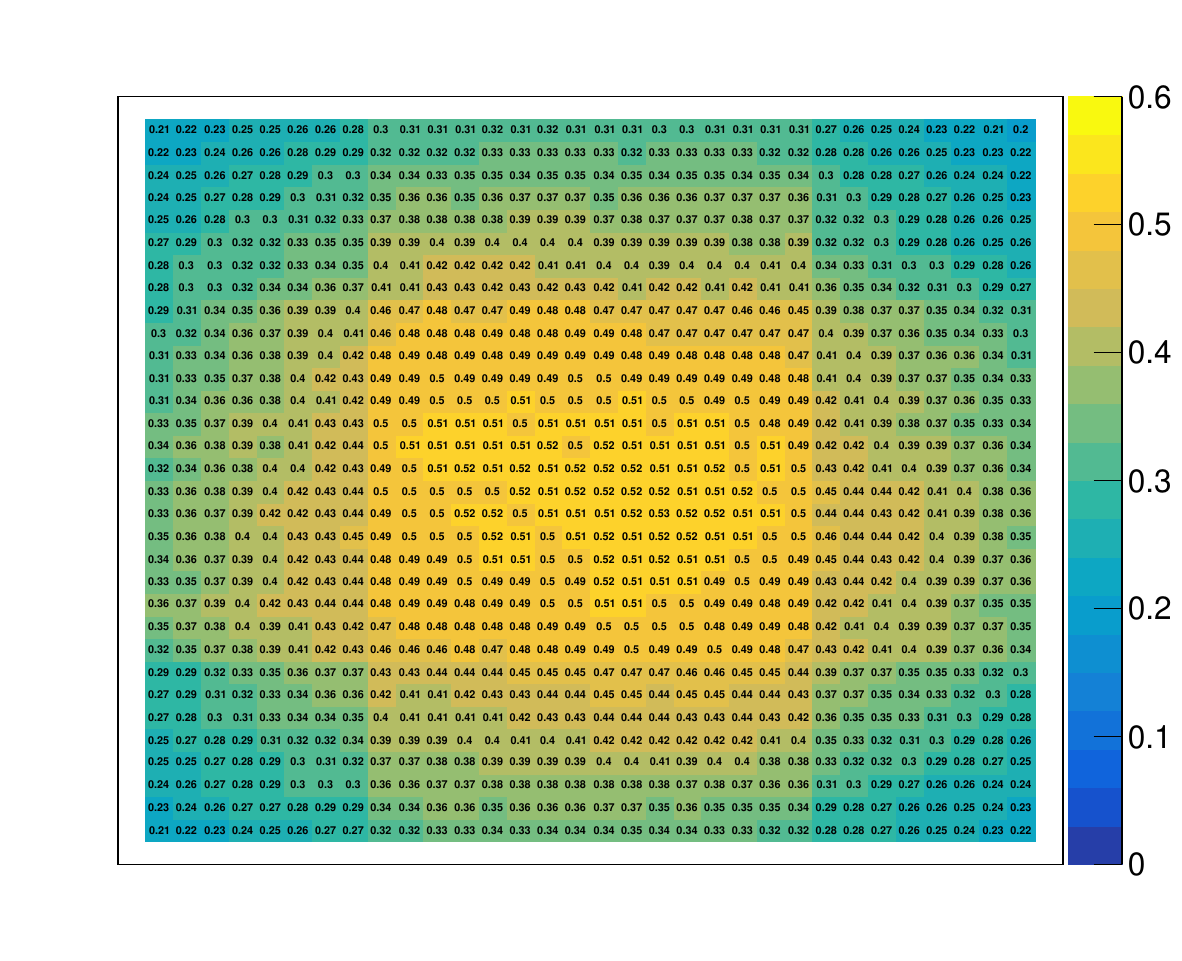} 
\caption{The light field map}
\label{fig:light-scan} 
\end{center}
\end{figure}

The light field scan is conducted using the reference SiPM, as discussed in Section~\ref{sec:setup}. The reference SiPM serves as a calibration standard for characterizing the light field across the SiPM tiles. It provides a known and stable response to incident light, allowing for accurate mapping of the light intensity distribution.
The data obtained from the light field scan is instrumental in calculating the absolute values of the PDE for the SiPMs on the individual tiles. 
It is visualized in Figure~\ref{fig:light-scan}, where the intensity levels are color-coded to depict the variations in light response.
Since the reference SiPM has a size of 6x6 mm², and the SiPMs on the tile are grouped in pairs measuring 12x12 mm², the numbers on the light field map should be summed up in groups of four.

\begin{figure}[htb]
\begin{center}  
\includegraphics[width=1.0\linewidth]{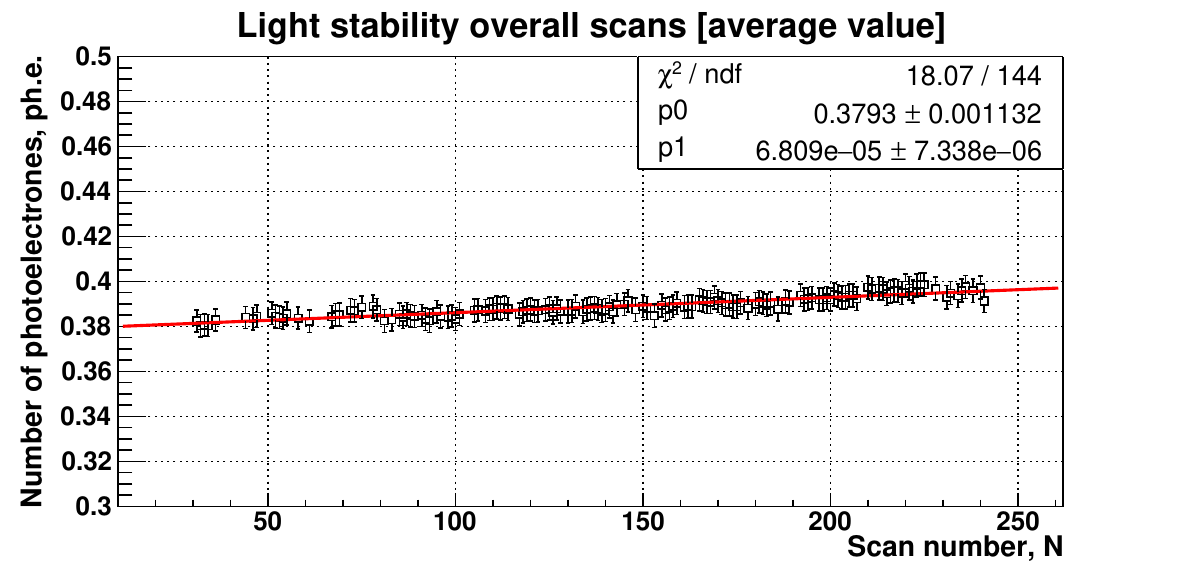} 
\caption{An example of the long term light stability over several light field runs.}
\label{fig:light-stability} 
\end{center}
\end{figure}

Another purpose is the estimation of the long-term stability of the light system. By analyzing the distribution over time, as shown in Figure~\ref{fig:light-stability}, it is possible to assess any temporal changes or fluctuations in the light field. This evaluation helps ensure the consistency and reliability of the SiPM response over extended periods.

\begin{figure}[htbp]
\centering
\includegraphics[width=.45\textwidth]{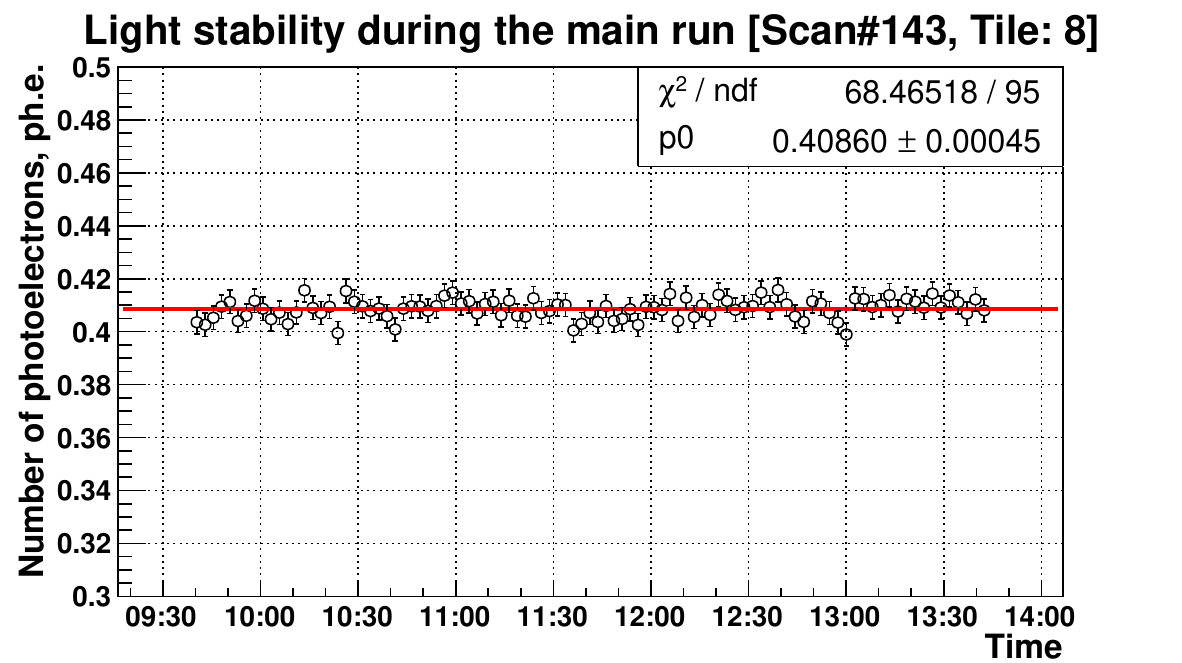}
\qquad
\includegraphics[width=.45\textwidth]{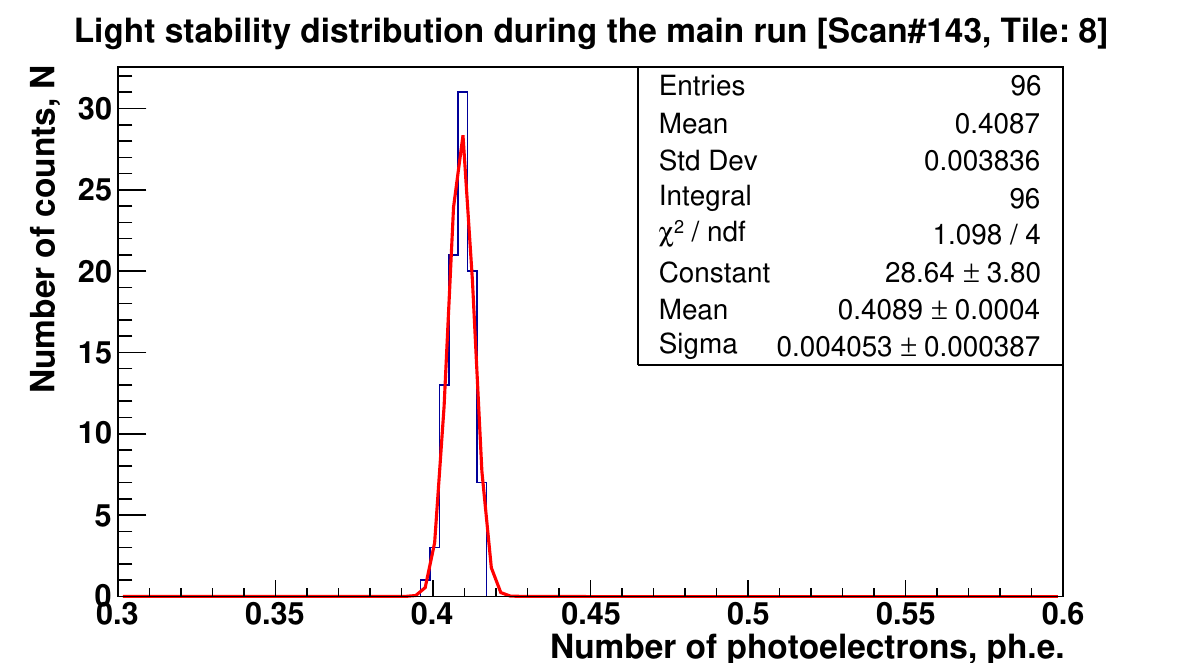}
\caption{(a) - Stability of the light intensity in time during the run 143 for tile 8 and (b) - the distribution of the light stability during the run 143 for tile 8. \label{fig:monitor-stability-tile-8}}
\end{figure}

During the main run, the information obtained from the reference SiPMs also enables the assessment of stability within the current run, i.e., monitoring. This stability analysis is depicted in Figure~\ref{fig:monitor-stability-tile-8}.  By continuously monitoring the response of the reference SiPMs throughout the run, any variations or fluctuations in local light intensity can be identified and analyzed.
In our measurements, we observe that the stability of the light intensity exhibits variations at a level of approximately 1\%. By  quantifying these fluctuations, we can account for any potential systematic effects and make appropriate adjustments to maintain the desired measurement precision.

\begin{figure}[htbp]
\centering
\includegraphics[width=.45\textwidth]{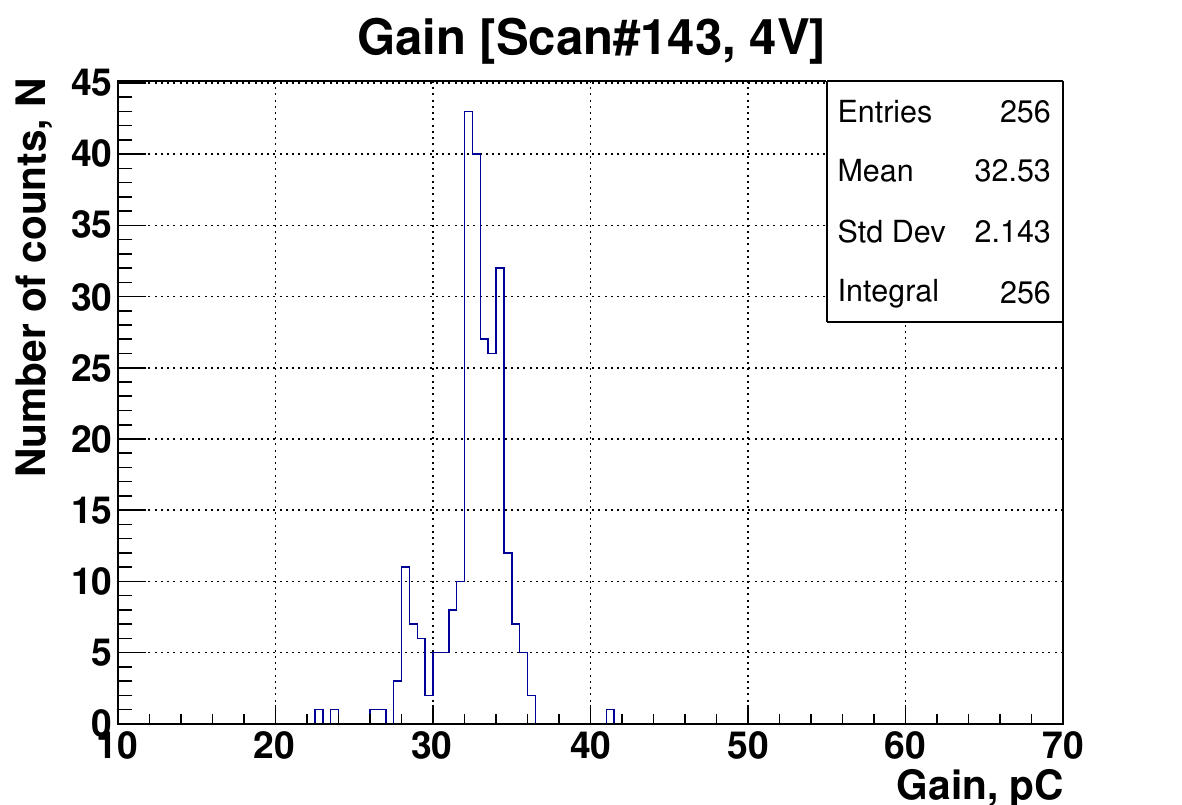}
\qquad
\includegraphics[width=.45\textwidth]{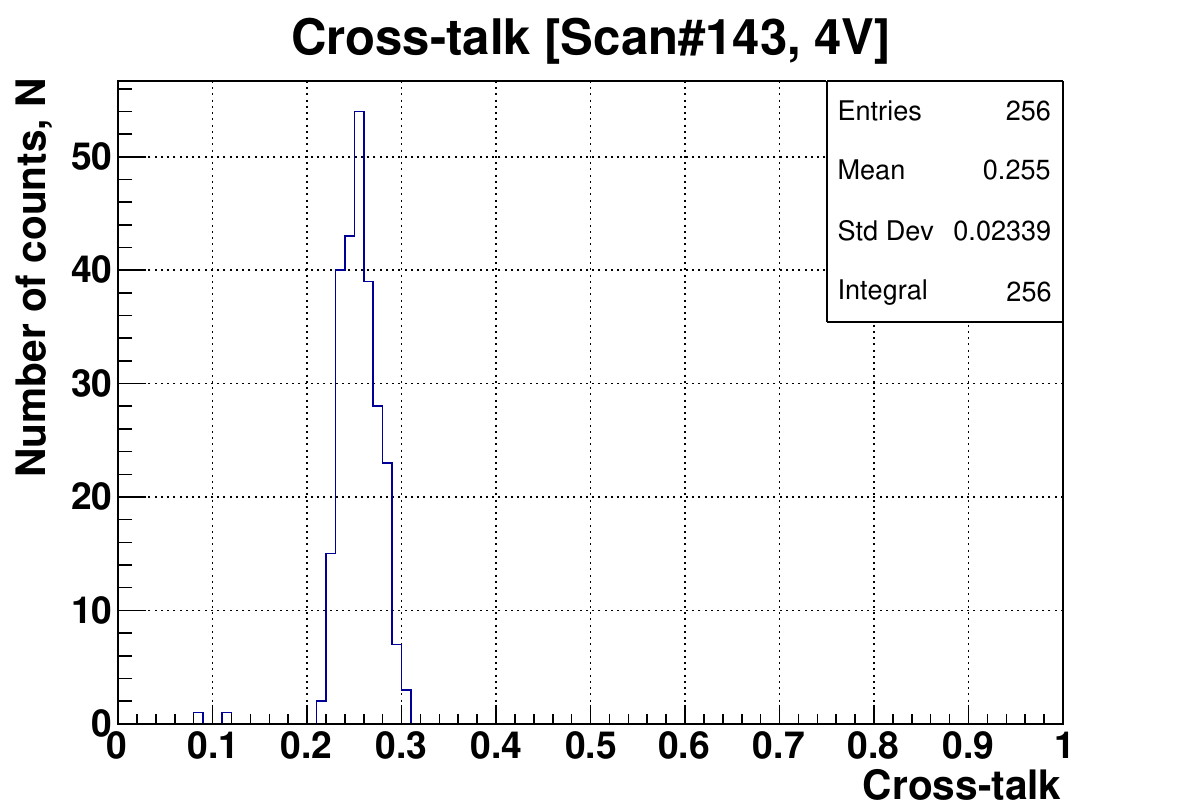}
\caption{An example of the distribution of values for 4 V overvoltage: (a) - gain, (b) - cross-talk}
\label{fig:gain_ct_dist}
\end{figure}

\begin{figure}[htbp]
\centering
\includegraphics[width=.45\textwidth]{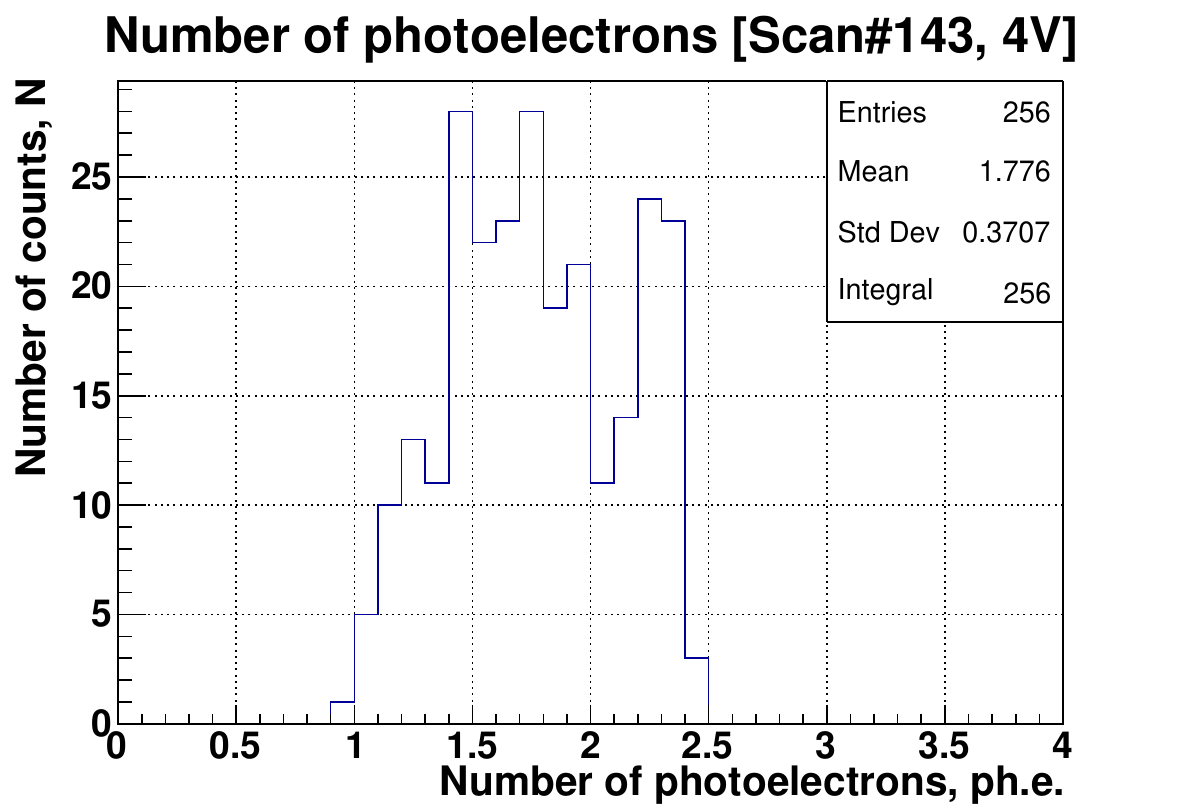}
\qquad
\includegraphics[width=.45\textwidth]{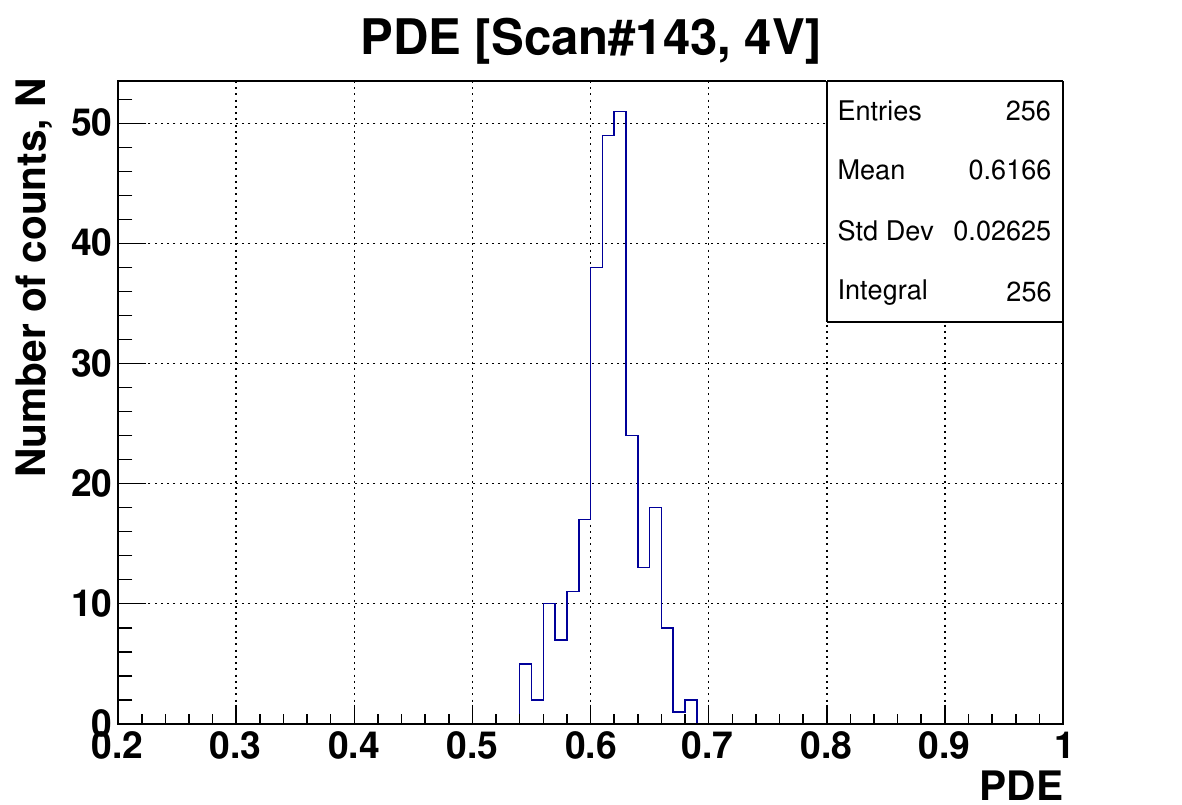}
\caption{An example of the distribution of values for 4 V overvoltage: (a) - number of photoelectrones, (b) - photon detection efficiency}
\label{fig:mu_pde_dist}
\end{figure}

\begin{figure}[htbp]
\centering
\includegraphics[width=.45\textwidth]{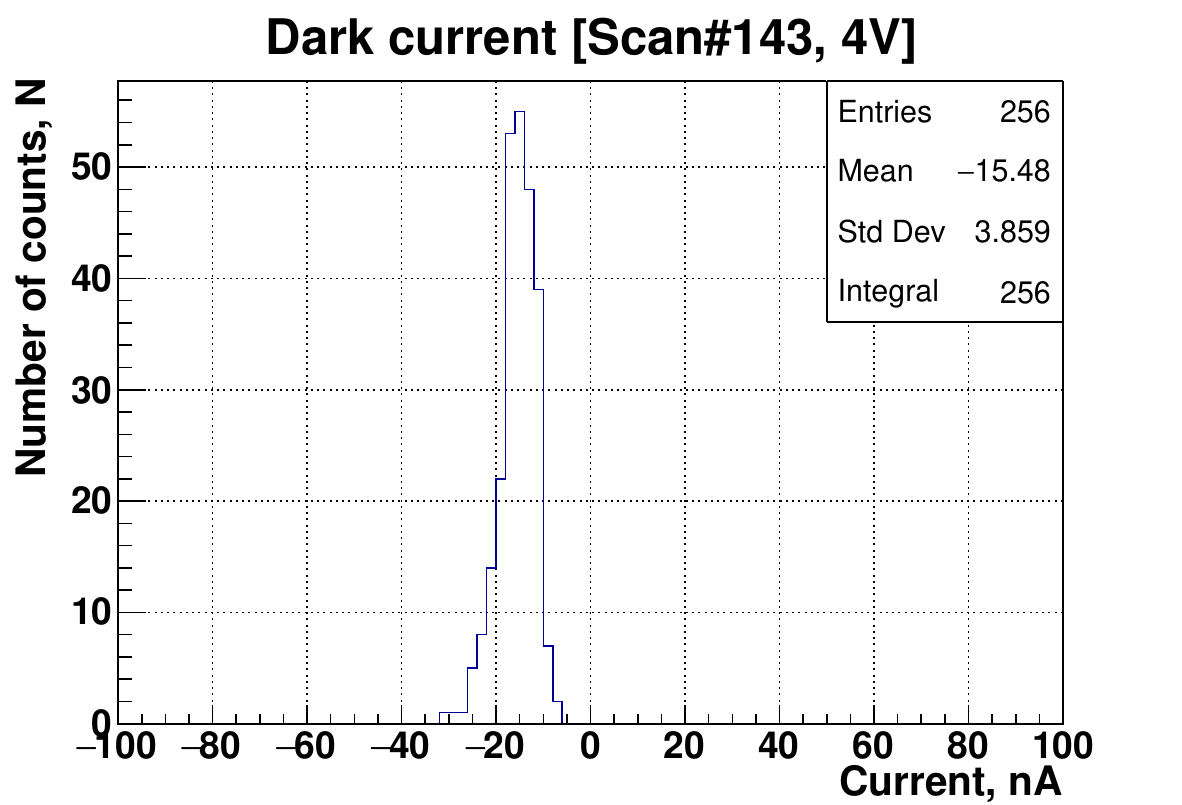}
\qquad
\includegraphics[width=.45\textwidth]{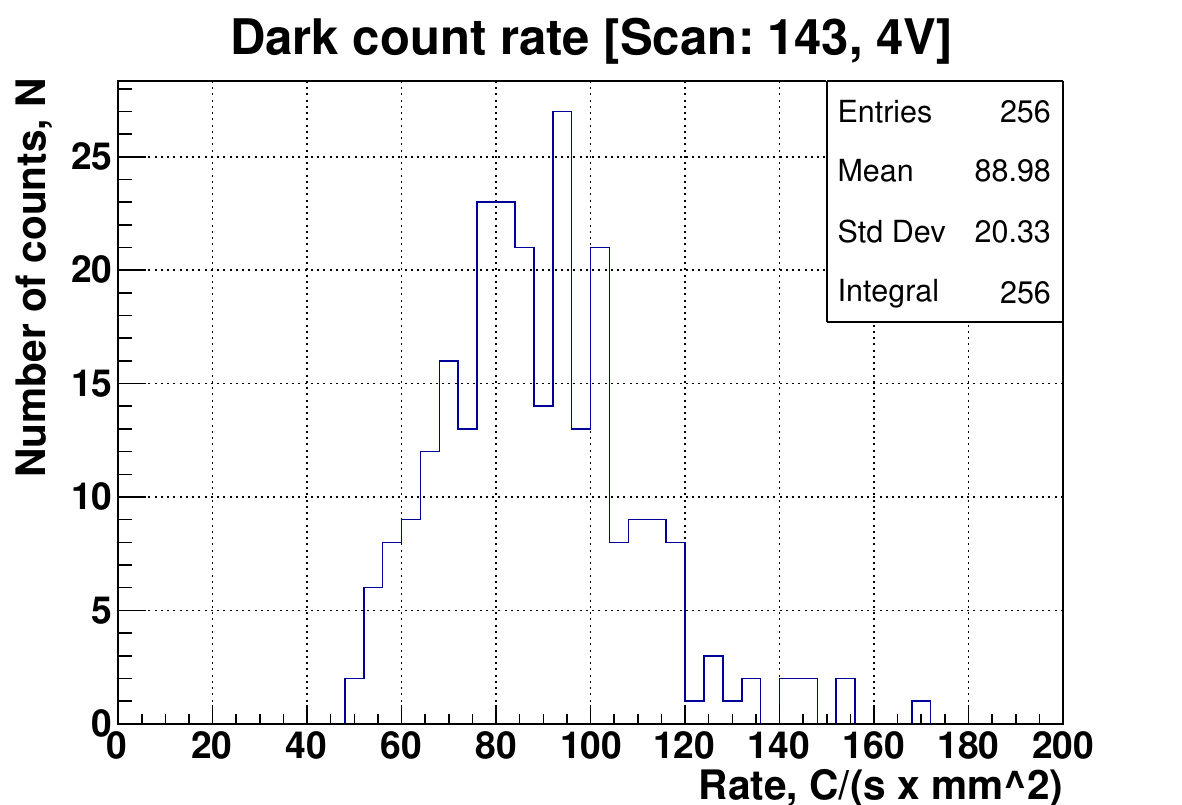}
\caption{An example of the distribution of values for 4 V overvoltage: (a) - dark current, (b) - dark count rate}
\label{fig:current_dist}
\end{figure}

The data files obtained during the main run are used to build charge spectra. The main parameters can be extracted from these spectra (see section 3), such as: gain, the number of photoelectrons and the probability of cross-talk for each of SiPM and different voltage points. DCR was estimated by means of the waveform analysis described in the same section. The dark current values for the same points were measured directly by means of the precised picoammeter. Examples of distributions of these parameters for one of the runs are shown in Figure ~\ref{fig:gain_ct_dist}, ~\ref{fig:mu_pde_dist} and ~\ref{fig:current_dist}. 

\begin{figure}[htb]
\begin{center}  
\includegraphics[width=.45\linewidth]{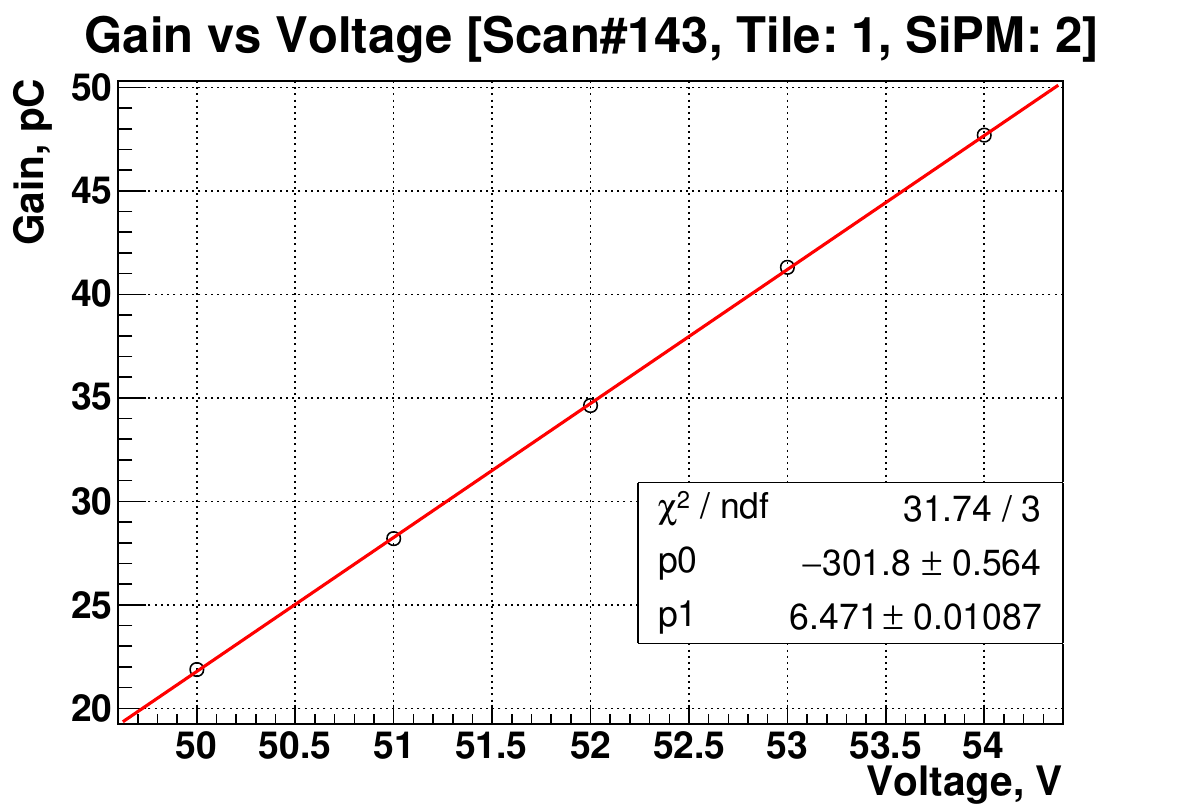} 
\qquad
\includegraphics[width=.45\textwidth]{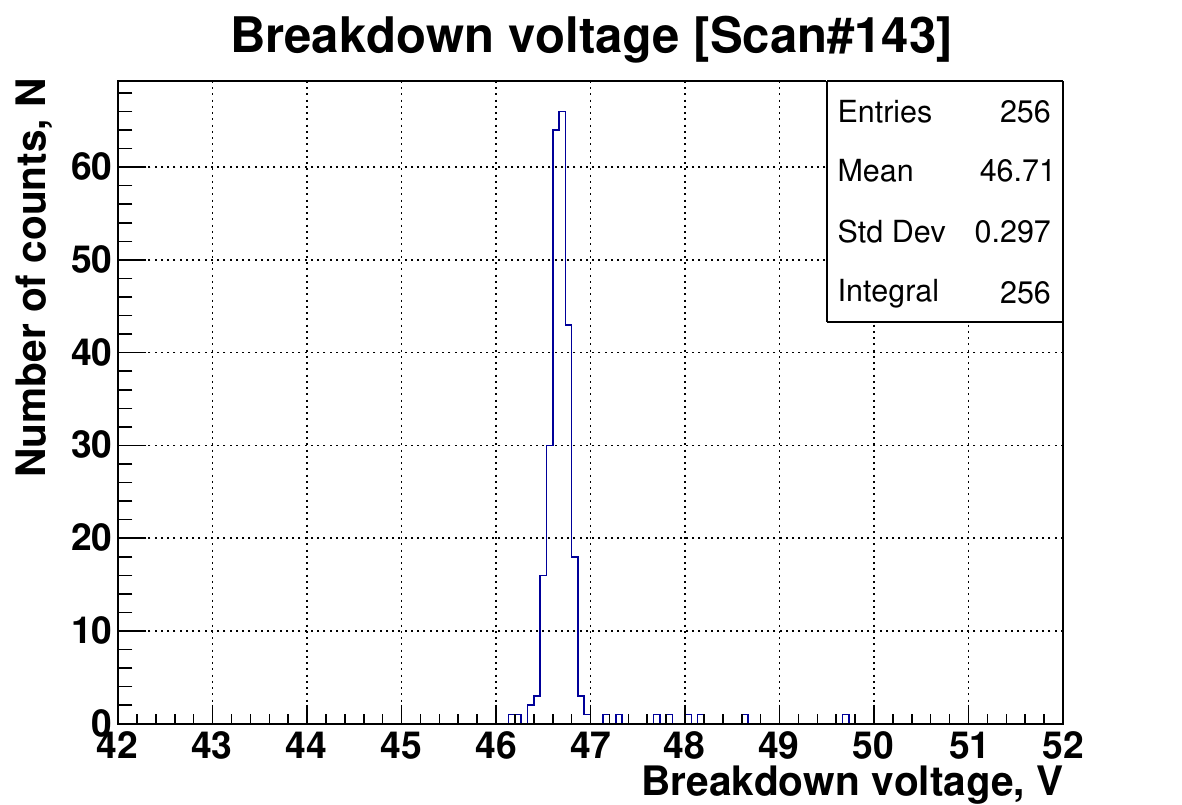}
\caption{(a) - Gain voltage dependence, (b) - breakdown voltage distribution overall SiPMs in a single run}
\label{fig:gain_vs_voltage} 
\end{center}
\end{figure}

The gain data obtained at different voltages allow us to find the dependences of the gain values on the voltage, and using linear regression to estimate the parameters of the linear function and calculate the breakdown points for each specific SiPM (figure~\ref{fig:gain_vs_voltage}a). Distribution of breakdown voltage for a particular run is presented on the Figure~\ref{fig:gain_vs_voltage}b.

\section{Summary}

This paper presents the design and methodology of a mass-testing setup utilized for the comprehensive characterization of SiPMs across all 4,100 tiles while operating at a negative temperature of -50°C. This setup enables the simultaneous testing of 16 tiles, equivalent to 256 SiPMs, in a single scan. It also facilitates the continuous monitoring of light intensity stability throughout the primary scan and the measurement of light field distribution over the SiPMs on each tile through the use of a translation stage. Control of the setup's equipment is managed via software implementation, which adopts a client-server approach. This approach offers the flexibility to integrate various devices into a unified system, ensuring convenient maintenance and scalability.

The data obtained using the monitor SiPMs indicate that the light intensity during the scanning of the SiPMs' tiles changes by less than 1\%. The light field scanning reveals a gradual increase in illumination at each scanning point, at a rate of approximately 1\% per month over time. This data can be employed for further refining the distribution of the light field during a specific main scan.

The examined parameters are derived from the charge spectra using the methods outlined in this paper. The distribution of these parameters demonstrates that the setup enabled us to observe variations in parameter values, allowing us to select tiles with similar characteristics. It allows voltage adjustments to align SiPM parameters, ensuring uniformity among SiPMs within the tile.

The methods developed in this research hold significant versatility and applicability, extending beyond the scope of the current study. They can be readily employed in a diverse range of applications for the mass-characterization of photodetectors under varying environmental conditions. This approach can be effectively adapted to assess not only for SiPM tiles but also arrays of small Photomultiplier Tubes (PMTs), multi-channel PMTs and batches of individual SiPMs. The robustness and adaptability of these methods make them valuable tools for ensuring the precise performance assessment of photodetector arrays across different scientific and industrial contexts.

\acknowledgments

This work is supported by the RSF-NSFC Cooperation Funding under Grant Numbers 21-42-00023 (RSF) and 12061131008 (NSFC).

\bibliographystyle{JHEP}
\bibliography{biblio.bib}

\end{document}